\documentclass[titlepage,12pt]{utarticle}
\usepackage{amsmath,amsfonts}
\long\def\omit#1{}

\numberwithin{equation}{section}

\newcommand{\homoquot}[3]{#3\Bigl\backslash\frac{#1}{#2}}
\newcommand{\mpd}[1]{M_{pl}^{\scriptscriptstyle (#1)}}

\begin{document}

\preprint{
UTTG--19--96\\
IASSNS--HEP--96/116\\
{\tt hep-th/9611104}\\
}

\title{(0,2) Noncritical Strings in Six Dimensions}

\author{Jacques Distler
 \thanks{ Work supported in part by NSF Grant PHY9511632,
  the Robert A.~Welch Foundation and an Alfred P.~Sloan Foundation Fellowship.}
 \address{ Theory Group, Physics Department\\ University of Texas at Austin\\
  Austin TX 78712 USA.  \\ {~}\\ \email{distler@golem.ph.utexas.edu}
  }
 and Amihay Hanany
 \thanks{Research supported in part by NSF Grant PHY-9513835.}
 \address{ Institute for Advanced Study\\ School of Natural Sciences\\
 Princeton, New Jersey\ 08540\\ {~}\\ \email{hanany@ias.edu}
 }
}

\date{November 14, 1996}

\Abstract{Type IIB strings compactified on K3 have a rich structure of
solitonic strings, transforming under $SO(21,5,\BZ)$.  We derive the BPS
tension formula for these strings, and discuss their properties, in
particular, the points in the moduli space where certain strings become
tensionless. By examining these tensionless string limits, we shed some
further light on the conjectured dual M-Theory description of this
compactification. }

\maketitle
\renewcommand{\baselinestretch}{1.25} \normalsize

\section{Introduction}

In the past two years we have witnessed a revolution in our understanding
of non-perturbative physics.
By looking closely at singular points in the moduli space of solutions to
string
vacua new fields massless degrees of freedom
 were discovered, which describe the local
dynamics of the singularity.
Some examples are Strominger's resolution of the conifold
\cite{Strominger:Conifolds},
and Witten's resolution of small $SO(32)$ instantons
\cite{Witten:SmallInstantons}.
In both of these cases the new massless states at the singularity are particles
which are either hypermultiplets - massless solitonic states, and/or vector
multiplets -- enhanced gauge symmetries.

On the other hand, the resolution of some singularities has involved the
introduction of new light degrees of freedom which are \emph{not}
particles. Witten's resolution of the singularities of type IIB
compactified on  K3  \cite{Witten:Strings95} involved the introduction of
tensionless strings. This is rather exotic infrared physics, of a sort that
had not been seen hitherto. Subsequently, tensionless strings have been
found to be responsible for nontrivial infrared physics in a variety of
string theory contexts \cite{Seiberg-Witten:Tensionless,Ganor-Hanany}. The
discovery of new nontrivial infrared physics is always exciting, especially
when it seems to be exotic from the field theory perspective. It is very
compelling to try to understand these tensionless string theories better.

In this note, we study the bound state spectrum of the BPS saturated
strings which arise in the original context of type IIB compactified on
K3. Upon further compactification to 5 dimensions, this theory becomes dual to
the heterotic  string on $T^5$, and we can exploit the known perturbative
behaviour of the heterotic string to learn some nonperturbative features of
this theory.


\section{$(0,2)$ Supersymmetry in 6 Dimensions}

As the vector of $SO(5,1)$ appears in the antisymmetric product of two
$\mathbf{4}$s, spinors in 6 dimensions can be taken to be
symplectic-Majorana-Weyl.  So the supercharges,$Q^a_\alpha$ of chiral $(0,N)$
supersymmetry carry both a spinor index, $\alpha$, and an $Sp(N)_R$ index, $a$.
For $N=2$, the case of interest in this note, the supersymmetry algebra is
\begin{equation}\label{eq:susyalg} 
\{Q^a_\alpha,Q^b_\beta\}=2\omega^{ab}
\gamma_{\alpha\beta}^\mu P_\mu+ \gamma_{\alpha\beta}^\mu Z^{ab}_\mu
\end{equation} where $\omega^{ab}$ is the $Sp(2)$-invariant tensor.
$Z^{ab}_\mu$ is a central charge of the supersymmetry algebra which transforms
as a Lorentz vector and as a $\mathbf{5}$ of $Sp(2)_R$.

As the central charge is a vector, the associated gauge field is a 2-form,
which naturally couples to a \emph{string}, rather than a \emph{particle}.

The massless representations of \eqref{eq:susyalg}  
(the central charge vanishes for particles) are
\begin{equation}\label{eq:reps}
\begin{split}
\text{gravity}\quad&(3,3;\mathbf{1})\oplus
(3,2;\mathbf{4})\oplus (3,1;\mathbf{5})\\
\text{tensor}\quad&(1,3;\mathbf{1})\oplus (1,2;\mathbf{4})\oplus
(1,1;\mathbf{5}) 
\end{split}
\end{equation}
where we denoted the representations of the little
group, $Spin(4)\, (=SU(2)\times SU(2))$ and $Sp(2)_R$. 

A 2-form has a 3-form field-strength which, in 6 dimensions, can be broken
into a self-dual and an anti-self-dual piece.  In light-cone gauge, these
correspond to self-dual 2-forms, $(3,1)$ of $Spin(4)$, and anti-self-dual
2-forms, $(1,3)$ of $Spin(4)$.  The former are part of the gravity multiplet;
the latter are part of tensor multiplets.  The central charge in
\eqref{eq:susyalg} measures the strength of the coupling of a string to the 5
self-dual 2-forms in the gravity multiplet (which transform as a $\mathbf{5}$
of $Sp(2)_R$).

Cancellation of gravitational anomalies require that the number of tensor
multiplets be equal to 21
\cite{AlvarezGaume-Witten:GravAnom,Townsend:K3Compactification}.  Each tensor
multiplet contains 5 real scalars, transforming as the $\mathbf{5}$ of
$Sp(2)_R$.  Consistent coupling to supergravity requires
\cite{Romans:6DSugra} that these 105 real scalars parametrize a manifold
which is, at least locally, of the form
\begin{equation*}
\frac{SO(21,5)}{SO(21)\times SO(5)}\quad .
\end{equation*}

The massless bosonic fields of the theory, thus, consist of the graviton, 5
self-dual 2-forms, 21 anti-self-dual 2-forms and these 105 scalars.  This is
precisely what emerges as the low-energy limit of Type II-B strings
compactified on $K3$.

The 2-forms arise as follows.  The NS-NS 2-form, $B$, and the R-R 2-form,
$\tilde B$, whose indices lie in the 6 noncompact directions each break up
into a self-dual and an anti-self-dual piece.  The $2^{nd}$ homology group,
$\homo{2}{K3}$ has signature (19,3).  Integrating the self-dual R-R 4-form
over a (anti-)self-dual 2-cycle on $K3$ yields an (anti-)self-dual 2-form in 6
dimensions.  All in all, we have the expected 26 2-forms:
\begin{equation}\label{eq:twoformsum}
\begin{split}
B^+,\,\tilde B^+,\,
\underset{\gamma\in\mathrm{H}^+_2(K3)}{\int G}\qquad&\text{5 self-dual 2-forms
in the gravity multiplet,}\\ 
B^-,\,\tilde B^-,\,
\underset{\gamma\in\mathrm{H}^-_2(K3)}{\int G} \qquad&\text{21 anti-self-dual
2-forms in the 21 tensor multiplets\quad .}
\end{split}
\end{equation}

Since there is a 26-dimensional space of 2-forms (21 anti-self-dual and 5
self-dual) to which a string might couple, we expect a rich spectrum of strings
in 6 dimensions.  Our aim is to find a BPS tension formula for these strings.

It is, perhaps, best to start by recalling the situation in 10 dimensions.
There there is a two-dimensional space of strings, which couple to some linear
combination of the two 2-forms, $B,\tilde B$.  The scalars take values in the
fundamental domain on the upper half-plane,
\begin{equation}
\tau=\tilde
\varphi + i \ex{-\varphi}\in \homoquot{SL(2,\BR)}{U(1)}{SL(2,\BZ)}
\end{equation} where
$\varphi$ is the dilaton and $\tilde\varphi$ is the R-R scalar.  The tension
of a string which couples to $n_1 B+n_2 \tilde B$ is given by
\cite{Schwarz:p-qstrings}
\begin{equation}\label{eq:pqtension}
T^2=\frac{1}{16\pi^2} \, (n_1\, n_2)\, \frac{1}{\Imag\tau}\begin{pmatrix}
1&-\Real\tau\\-\Real\tau&\lvert\tau\rvert^2 \end{pmatrix}
\begin{pmatrix} n_1\\
n_2 \end{pmatrix}\, (\mpd{10})^4\quad .
\end{equation}
This is the natural
$SL(2,\BZ)$-invariant expression, where $SL(2,\BZ)$ takes
\begin{gather}
\tau\to
\frac{a\tau+b}{c\tau+d}\notag\\ 
\begin{pmatrix}B\\ \tilde B\end{pmatrix} \to
\begin{pmatrix}dB- b\tilde B\\ -cB+ a\tilde B\end{pmatrix},\qquad
\begin{pmatrix}n_1\\ n_2\end{pmatrix} \to \begin{pmatrix}a n_1 +b n_2\\c n_1 +d
n_2\end{pmatrix}\quad .
\end{gather}

In addition to this set of strings which couple to $B$ and $\tilde B$, new
strings arise in 6 dimensions from wrapping the Type IIB 3-brane around
2-cycles on the $K3$.  Since the 3-brane couples to the self-dual 4-form,
$G$, these strings couple to the 2-forms we obtained by integrating $G$ over
the corresponding 2-cycle on $K3$.  Once we turn on background fields, these,
apparently distinct, sets of strings mix in a complicated way.  Our task is to
write down the analogue of \eqref{eq:pqtension}.

The results are as follows.  The space of scalars
\cite{Aspinwall:U-Duality}
\begin{equation}\label{eq:modspace} \CM=\homoquot{SO(21,5)}{SO(21)\times
SO(5)}{\Gamma} \end{equation}
is the moduli space of \emph{even, self-dual
Lorentzian lattices of signature $(21,5)$} (21 negative and 5 positive
eigenvalues).  The discrete group $\Gamma$, usually written as $SO(21,5,\BZ)$,
is the discrete subgroup of $SO(21,5)$ which acts as automorphisms of some
chosen basepoint lattice, $\Lambda_0$.  Let $\{e_i\}$ be a basis for
$\Lambda_0$.  Then, if $G(\phi)$ is an $SO(21,5)$ transformation which takes us
from the basepoint to the point $\phi\in\CM$,
$\boldsymbol{E}_i(\phi)=G(\phi)\boldsymbol{e}_i$ is a basis for the
corresponding lattice, $\Lambda_\phi$.

The Dirac quantization condition on the strings simply states that the allowed
charges, under the (21,5)-dimensional space of 2-forms, lie on this even,
self-dual Lorentzian lattice.  That is, an allowed charge vector is an integer
linear combination $\sum_i n_i \boldsymbol{e}_i$.  Of course, the description
in terms of a fixed basis of 2-forms is good only locally in the moduli space.
Under the action of the modular group, $\Gamma$, the 2-forms transform as the
fundamental 26-dimensional representation.  So it is more convenient for our
purposes to work with the ``charge vector" $\boldsymbol{q}(\phi)=\sum_i n_i
\boldsymbol{E}_i(\phi)$, a section of a flat vector bundle over $\CM$ with
structure group $\Gamma$.

The condition for a string to be BPS-saturated requires first that
\begin{equation}\label{eq:lengthsquared}
\boldsymbol{q}^2\geq -2\quad .
\end{equation}

Introduce the orthogonal projection, $P$, which projects any vector onto the
5-dimensional positive-signature subspace. Define
\begin{equation}
\boldsymbol{q}_+=P\boldsymbol{q},\qquad
\boldsymbol{q}_-=(\Bid-P)\boldsymbol{q}\quad .
\end{equation}
In particular, $\boldsymbol{q}_+$ simply measures the strength of the
coupling  of the string to the 5 self-dual 2-forms in the gravity multiplet.
The BPS tension formula is
\begin{equation}\label{eq:BPStension}
T^2 = \frac{1}{8\pi^2} \lvert
\boldsymbol{q}_+(\phi)\rvert^2\, (\mpd{6})^4
\end{equation}
where, here, $\mpd{6}$ is
the \emph{6-dimensional} Planck mass. 

Consider a BPS-saturated string with charge-vector
$\boldsymbol{q}=\boldsymbol{q}_1+\boldsymbol{q}_2$ equal to the sum of the
charge-vector of two other BPS-saturated strings.  This string is stable
against decaying into the two other strings provided
\begin{equation}\label{eq:stable}
T< T_1+T_2\quad .
\end{equation}
By the
triangle-inequality and \eqref{eq:BPStension}, we always have
\begin{equation*}
T\leq T_1+T_2
\end{equation*}
with equality only for
$(\boldsymbol{q}_{1})_+ = c(\boldsymbol{q}_{2})_+ $ for some nonnegative
constant $c$.

\subsection{The fundamental string}\label{sec:fundstring}

Where is the fundamental Type IIB string?  Consider a subspace of $\CM$ on
which there is a distinguished even self-dual sublattice of dimension $(1,1)$,
corresponding to the anti-self-dual and self-dual components of $B$.  There
are two strings corresponding to the basis vectors of this lattice.  The
first, corresponding to the basis vector
$\boldsymbol{e}_1=\tfrac{1}{\sqrt{2}}(1,1)$, couples to the sum
$(B^-+B^+)=B$, and is the fundamental Type IIB string.  The second,
corresponding to the basis vector
$\boldsymbol{e}_2=\tfrac{1}{\sqrt{2}}(1,-1)$, couples to the difference
$(B^--B^+)$ and is solitonic.  There is a 1-parameter family of such
distinguished (even, but not self-dual) sublattices, corresponding to
\begin{equation}\label{eq:subfam}
\begin{split}
\boldsymbol{E}_1&=\ex{\varphi}\boldsymbol{e}_1\\
\boldsymbol{E}_2&=\ex{-\varphi}\boldsymbol{e}_2\quad .
\end{split}
\end{equation}
The tensions of
these strings are \eqref{eq:BPStension}
\begin{equation}\label{eq:twostrings}
T_1^2=\frac{1}{16\pi^2}\ex{2\varphi} (\mpd{6})^4,\qquad
T_2^2=\frac{1}{16\pi^2}\ex{-2\varphi} (\mpd{6})^4\quad .
\end{equation}
Recalling that, in $d$ dimensions,
\begin{equation}\label{eq:MplMs}
\mpd{d}=\ex{-\frac{2}{d-2}\varphi} M_s
\end{equation}
we see that, as expected, the former has a tension equal to
$\frac{M_s^2}{4\pi}=\frac{1}{2\pi\alpha'}$, whereas the latter is very heavy
 at weak coupling ($T_2=\frac{\ex{-2\varphi}}{2\pi\alpha'}$).

The string with charge vector
\begin{equation*}
\boldsymbol{q}=\boldsymbol{E}_1+\boldsymbol{E}_2=\sqrt{2}
(\cosh\varphi,\sinh\varphi)
\end{equation*}
couples to the anti-self-dual part
of $B$.  At $\varphi=0$, the tension of this string
\begin{equation*}
T=\frac{1}{2\pi}\ (\mpd{6})^2\ \lvert \sinh\varphi\rvert
\end{equation*}
vanishes.
At this point the fundamental string and the string which couples to
$(B^--B^+)$
become degenerate.

\subsection{Tensionless strings}

Obviously, if $\boldsymbol{q}_+=0$, then the tension of the corresponding
string vanishes.  Of course, $\boldsymbol{q}$ cannot be identically $0$; the
string must have some charge.  The condition \eqref{eq:lengthsquared}
$q^2\geq-2$ then implies that such a string has
\begin{equation}\label{eq:tensionless}
q^2=-2,\qquad \boldsymbol{q}_+=0.
\end{equation}
The set of such vectors spans an even (negative-definite)
Euclidean sublattice of $\Lambda_\phi$.  In fact, such a set of vectors forms
the root system for a simply-laced (since all the vectors have
(length)$^2=-2$) Lie algebra.

When do such tensionless strings arise?  Some are associated with
singularities of the $K3$ surface; when certain 2-cycles on the $K3$ shrink,
the corresponding wrapped 3-branes become tensionless.  Such singularities
naturally have an ADE classification, and we get tensionless strings
associated to the corresponding ADE root system.

Similarly, as we have seen, the string which couples to the anti-self-dual part
of $B$ becomes tensionless at $\varphi=0$ (and suitable values for the other
background fields, in particular, $\tilde\varphi=0$).

\section{Relation to the Heterotic String in 5 Dimensions}

We compactify from 6 down to 5 dimensions on a circle of radius $r$ (in the
Einstein metric).  A string which wraps $n$ times around the circle
corresponds to a particle of mass
\begin{equation}\label{eq:wrap}
m =2\pi\lvert n\rvert r T\quad .
\end{equation}
We use T-duality to relate this to the Type IIA string
compactified on $K3\times S^1$ and then use string-string duality to relate
this to the heterotic string compactified on $T^5$.

Here we recognize
\begin{equation*}
\CM=\homoquot{SO(21,5)}{SO(21)\times
SO(5)}{\Gamma}
\end{equation*}
as the Narain moduli space
\cite{Narain:toroidal,Narain-Sarmadi-Witten} of heterotic strings compactified
on $T^5$.  What we called the charge vector $\boldsymbol{q}$ previously is
simply the internal momentum of the toroidally-compactified heterotic string.
\begin{equation}\label{eq:qcorresp}
\boldsymbol{q}_R= \boldsymbol{q}_+,\qquad
\boldsymbol{q}_L=\boldsymbol{q}_-\quad .
\end{equation}
The heterotic string
coupling is given in terms of the radius of the circle by
\begin{equation}\label{eq:hetcoupling}
\ex{2\varphi_H/3}=\frac{1}{\sqrt{2}}
\mpd{5} r\quad .
\end{equation}

The (perturbative) mass formula for a heterotic string state with internal
momentum $\boldsymbol{q}=(\boldsymbol{q}_L,\boldsymbol{q}_R)$ is (in units of
the heterotic string scale)
\begin{equation}\label{eq:hetmass}
\begin{split}
m^2&= q_L^2+2(N_L-1)\\ 
&=q_R^2 +2N_R\quad .
\end{split}
\end{equation}
A BPS-saturated state (the only
thing we have any right to compare with the Type II string) has $N_R=0$.  So,
for BPS-saturated states, $m^2=q_R^2$ and
\begin{equation}
q_R^2-q_L^2=2N_L-2\geq -2\quad .
\end{equation}

Putting \eqref{eq:wrap}, \eqref{eq:hetcoupling} and \eqref{eq:hetmass}
together,  and using the relation
\begin{equation*}
(\mpd{5})^3= r (\mpd{6})^4
\end{equation*}
we determine the tension of the putative wrapped strings.

Of course, in outline, the proof of the tension formula
\eqref{eq:BPStension} does not really rely at all on the details of the chain
of dualities we have used. Once we realize that, upon compactification on a
circle of radius $r$, a wrapped string corresponds to a particle of mass $M=
2\pi r T$, we can obtain the BPS bound on the masses of these particles
simply by manipulating the dimensionally-reduced supersymmetry algebra
\eqref{eq:susyalg}. The little group for this situation is
$Spin(4)=SU(2)\times SU(2)$. The supercharges of \eqref{eq:susyalg}
transform as doublets of one of the $SU(2)$s and can be written as $Q^a_i,
Q^{\dagger i}_a$, where $i$ is an $SU(2)$ index. The supersymmetry algebra
becomes
\begin{equation}\label{eq:ReducedSusyAlg}
\begin{split}
\{Q^a_i,Q^{\dagger j}_b\}&=2\delta^a_b\delta_i^j M\\
\{Q^a_i,Q^b_j\}&=\epsilon_{ij} z^{ab}\\
\{Q^{\dagger i}_a,Q^{\dagger j}_b\}&=\epsilon^{ij}z_{ab}
\end{split}
\end{equation}
where $z^{ab}= \int_{S^1} Z^{ab}_\mu dx^\mu$ and we raise and lower $Sp(2)$
indices using $\omega_{ab}$. The standard trick is to consider now the
operator
\begin{equation*}
A^a_i=Q^a_i-\frac{1}{2M} \epsilon_{ij}z^{ab}Q^{\dagger j}_b\quad .
\end{equation*}
We then have the positivity condition
\begin{equation}
\begin{split}
0\leq& \{A^a_i,A^{\dagger j}_a\}\\
=& \delta^j_i \bigl[ 8M -\frac{1}{2M} z_{ab}z^{ab}\bigr]
\end{split}
\end{equation}
from which we conclude $M^2\geq \frac{1}{16}\lvert z\rvert^2$.

On the other hand, the connection with the heterotic string gives us some
physical information which we can now exploit to argue for the
existence and uniqueness of bound states of strings for a given charge vector
$\boldsymbol{q}=\sum n_i
\boldsymbol{E}_i$. The corresponding string is stable \emph{if and only if}
the $\{n_i\}$ have no common factor.

This follows from comparing the spectrum of wrapped strings to
the perturbative spectrum of BPS states of the heterotic string on $T^5$.
Given a string with charge vector $\boldsymbol{q}$, if there were a stable
string with charge vector $k\boldsymbol{q}$, then there would be, upon
compactifying down to 5 dimensions, two way to obtain a BPS state with charge
vector
$k\boldsymbol{q}$ in 5 dimensions: we could wrap the $\boldsymbol{q}$-string
\emph{$k$ times} around the circle, or we could wrap the $k\boldsymbol{q}$
\emph{once}. But, in the heterotic theory, there is only one BPS state with
charge vector $k\boldsymbol{q}$. So, since the $k$-times wrapped
$\boldsymbol{q}$-string \emph{must} exist, the $k\boldsymbol{q}$ must not
exist as a stable BPS state. Exactly the same argument can be used to rule
out bound states at threshold for the $(p,q)$ strings in 10 dimensions
\cite{Schwarz:p-qstrings}.

At a generic point in the moduli space, the gauge group in 5 dimensions is
$U(1)^{27}$.  The bosonic part of the supergravity Lagrangian is
\cite{Awada:5Dsugra}
\begin{multline}\label{eq:sugraL}
\CL= (\mpd{5})^3\sqrt{-g}\left[
-\frac{1}{2}R-\frac{1}{4(\mpd{5})^2}\ex{2\varphi_H/3}a_{ij}F^i\cdot F^j -
\frac{1}{4(\mpd{5})^2}\ex{-4\varphi_H/3} F^2\right.\\
\left. -\frac{1}{6}(\partial\varphi_H)^2
-\frac{1}{2}\gamma_{\alpha\beta}\partial\phi^\alpha\partial\phi^\beta
\right] +\frac{\sqrt{2}}{8} C_{ij} F^i\wedge F^j\wedge A\quad .
\end{multline}
Here, as before, the $\phi$ parametrize the moduli space, $\CM$, and
$\gamma_{\alpha\beta}$ is the $SO(21,5)$-invariant metric.  $\varphi_H$ is a
scalar in the gravity multiplet, which is either the heterotic string dilaton
or, using \eqref{eq:hetcoupling}, the dilatation mode of the circle in the IIB
picture. $A$ is the graviphoton (with field strength
$F=dA$), $A^i$ (with field strengths $F^i=dA^i$) are 26 $U(1)$ gauge fields,
21 of which are part of 5-D vector multiplets, and 5 of which transform as the
$\boldsymbol{5}$ of $Sp(2)$ and are part of the gravity multiplet.  $C_{ij}$
is the constant matrix
\begin{equation}
C_{ij}=E_i(\phi)\cdot
E_j(\phi)=e_i\cdot e_j \end{equation} and \begin{equation}\label{eq:gaugekin}
a_{ij}=2 E_i(\phi)_+\cdot E_j(\phi)_+ - C_{ij}\quad .
\end{equation}

In the heterotic string picture, the $U(1)^{26}$ is simply the unbroken gauge
group that one has at a generic point in the Narain moduli space. The
$27^{th}$ $U(1)$, the graviphoton, arises from the antisymmetric tensor field,
$B_{\mu\nu}$, which is dual to a 1-form in 5 dimensions. Among the
$U(1)^{26}$ gauge bosons, 5 are ``right-moving", while the rest are
``left-moving". The  ``right-moving" gauge bosons transform in the
$\boldsymbol{5}$ of $Sp(2)$, and are part of the gravity supermultiplet.

In the IIB picture, the graviphoton arises as the Kaluza-Klein reduction of
the 6-dimensional metric, compactified on $S^1$. The other 26 gauge fields
arise as the reductions of the 2-forms that were present in 6 dimensions.
(Since we decomposed the 2-forms in 6 dimensions into their self- and
anti-self-dual pieces we can, without loss of generality, take one of the
indices of each to be tangent to the $S^1$.) Again, 5 correspond to self-dual
and 21 to anti-self-dual 2-forms.

At special points in the moduli space, the $U(1)^{21}$ symmetry (excluding
the graviphoton, and the 5 ``right-moving" photons in the gravity multiplet)
gets enhanced to a nonabelian gauge symmetry. In the heterotic picture, this
occurs when there are vectors in the Narain lattice with
$\boldsymbol{q}_R=0$,
$q_L^2=2$. But, given the correspondence \eqref{eq:qcorresp}, we
see that this is precisely where the IIB theory develops tensionless strings
\eqref{eq:tensionless}. The cartan generators correspond, as we have seen, to
anti-self-dual 2-forms in 6 dimensions. The rest of the generators correspond
to tensionless strings, wrapped around the $S^1$.

Using \eqref{eq:hetcoupling},\eqref{eq:MplMs}, we see that the inverse gauge
couplings,
\begin{equation*}
\left(\frac{1}{g^2}\right)_{ij}=\mpd{5}\ex{2\varphi_H/3}a_{ij}=
M_{s} a_{ij}
\end{equation*}
are given by essentially the same expression as
we found above, \eqref{eq:wrap},\eqref{eq:BPStension}, for the masses of the
wrapped strings.  The graviphoton inverse gauge coupling is
\begin{equation*}
\frac{1}{g^2}=\mpd{5}\ex{-4\varphi_H/3}=\frac{2^{3/4}}{M_{s}^{1/2} r^{3/2}}.
\end{equation*}
which becomes weakly-coupled as we take the radius $r\to0$.

The gauge coupling near the $A_1$ singularity discussed in section
\ref{sec:fundstring} is
\begin{equation*}
\left(\frac{1}{g^2}\right)_{ij}=\frac{1}{\sqrt{2}} (\mpd{5})^2 r
\begin{pmatrix}\ex{2\varphi}&0\\0&\ex{-2\varphi}\end{pmatrix}
\end{equation*}
and so, near the singularity we have two strongly coupled fields which
degenerate while away
from the singular point one becomes weakly coupled and the other strongly
coupled. Note that near the singularity we have a $Z_2$ symmetry which
exchanges
the two entries in the matrix. This is a typical remnant of an $A_1$
singularity.
Up on the covering space, $SO(21,5)/SO(21)\times SO(5)$, the gauge
couplings are smooth near the phase transition points associated to
tensionless strings. However, when we mod out by the discrete symmetry
$SO(21,5,\BZ)$, these transition points are orbifold singularities in $\CM$.

One na\"\i vely might wonder about the tensionless strings which are
\emph{not wrapped} around the $S^1$, but which propagate in the 5-dimensional
spacetime. Since, in 5 dimensions, a string is dual to a particle, these
strings are
\emph{magnetic} sources for the gauge fields we have been studying. For small
$r$, they are very heavy compared to the typical particle mass
\eqref{eq:wrap}. So, even in the limit where both are going to zero (that is,
when the underlying string in 6 dimensions is becoming tensionless), the
particle, being so much lighter, dominates the infrared physics.

\subsection{Strong heterotic coupling}

One of the most remarkable developments in this field was the observation by
Witten \cite{Witten:variousdims} and Townsend \cite{Townsend:revisited} that
the strong coupling limit of type IIA string theory in 10 dimensions is a
theory with 11-dimensional Lorentz invariance.  Here we see that, as a
consequence of
\eqref{eq:hetcoupling}, the strong coupling limit of the heterotic string in 5
dimensions (compactified on $T^5$) is a theory with 6-dimensional
Lorentz-invariance, the type IIB string compactified on $K3$!

Of course, in this limit of large heterotic coupling, or large $r$, the
situation described in the last paragraph of the previous subsection is
turned on its head. In this case, it is the magnetically-charged
\emph{strings} which are much lighter than the electrically-charged
\emph{particles}. It is the \emph{strings} which dominate the infrared
physics\footnote{Of course, our ability to make these statements, in the
limit where \eqref{eq:sugraL} becomes strongly-coupled, relies on the fact
that the N=4 supersymmetry forbids corrections to the BPS formul\ae\ for
these masses.}.

The BPS saturated particle multiplets of theories with enough supersymmetry
can
be classified into
     \emph{short} (annihilated by half of the supersymmetries)
and
     \emph{ultrashort} (annihilated by 3/4 of the supersymmetries)
multiplets. In the case of toroidally-compactified heterotic strings,
these correspond to
\begin{description}
\item[short]$N_R=0$
\item[ultrashort]$N_L=N_R=0$
\end{description}
We saw that those BPS strings which can become tensionless always have
$q^2=-2$, {\it i.e.}, they
correspond to ultrashort multiplets upon compactification. The strings (like
the fundamental string) with $q^2\geq 0$ which never become tensionless
correspond to short multiplets.
The concept of short and long multiplets certainly makes sense for the
dimensionally-reduced supersymmetry algebra \eqref{eq:ReducedSusyAlg}.
Perhaps a similar concept can be made sense of for
\emph{strings}  in the 6 dimensional (0,2) supersymmetry algebra
\eqref{eq:susyalg}.

\section{M-Theory Picture}

It was shown in \cite{Mukhi:IIBdual,Witten:IIBdual} that the dual theory to
type IIB on $K3$ is M-theory on a $T^5/\BZ_2$.  The $\BZ_2$ acts as $-1$ on
all circles of the torus.  It also acts on the 3-form gauge field as
$-1$.  This symmetry breaks half of the supersymmetries and leaves all
generators which obey
$\Gamma_{7,8,9,10,11}\epsilon=\epsilon$. This theory gives chiral
supersymmetry in six dimensions, namely the $(0,2)$ supersymmetry discussed
above.  The condition on the generators is consistent with having 5-branes
which are localized in the
$7,8,9,10,11$ directions. Actually an anomaly cancellation argument leads to
having $16$ five branes.  It was stressed in
\cite{Witten:IIBdual} that gravitational anomalies must cancel locally in
spacetime.  This led to the observation that the $32$ fixed points on the
torus carry magnetic charge
$-\frac{1}{2}$ (in the units where each of the 16 five-branes carries magnetic
charge $+1$).  In addition to cancelling the anomalies, this satisfies
magnetic charge conservation.

The untwisted sector contains  $5$ 2-form fields, given by integrating the
3-form of M-theory over a 1-cycle in $T^5$. Since both the 3-form and the
1-cycle are odd under the $\BZ_2$, these states are even, and survive the
orbifold projection.  The self-dual parts of these 2-forms correspond to the
$5$ self-dual 2-forms in the
$(0,2)$ gravity multiplet.  The anti-self-dual parts give rise to 5 tensor
multiplets. In addition we have $10$ scalars from the 3-form field (with
all indices tangent to $T^5$) and $15$ scalars from the modes of the metric
on $T^5$.  Together with the $5$ anti-self-dual 2-form fields they form
the  $5$ tensor multiplets.  These are the massless fields in the untwisted
sector.

In addition to the tensor multiplets from the untwisted sector, there are 16
more tensor multiplets, one carried by each of the 5-branes. Each tensor
multiplet has an anti-self-dual 2-form and five scalars, which give the
position of the 5-brane on $T^5/\BZ_2$.

The M-theory description of the 105-dimensional moduli space clearly has 25
untwisted moduli related to the geometry of the $T^5$ and 80 twisted moduli
related to the positions of the 5-branes. It is tempting to try to identify
the former with the moduli of the torus in the compactification
of the heterotic string on $T^5$ and the latter with the Wilson lines of the
heterotic compactification. This is almost, but not quite, correct. 

Let $X$ be the $T^5$ of the M-theory compactification. The space of twisted
moduli is
\begin{equation}\label{eq:TwistedMod}
X^{16} / \Gamma
\end{equation}
where $\Gamma$ is the group $S_{16}\semidir(\BZ_2)^{16}$ with generators
\begin{equation}\label{eq:GammaGen}
\begin{split}
\sigma_{ij}:&\ \vec x_i \leftrightarrow \vec x_j \\
s_i:&\ \vec x_i\to -\vec x_i\quad .
\end{split}
\end{equation}
The reason for modding out by $\Gamma$ is evident. Since we do not label the
5-branes, we should mod out by transformations which permute them. Also,
since the 5-branes are really propagating on $X/\BZ_2$, not $X$, we should
consider $\vec x_i$ and $-\vec x_i$ as equivalent.

Let us compare this with the space of Wilson lines of the $Spin(32)/\BZ_2$
heterotic string on $Y=T^5=\BR^5/\Lambda$. This is
\begin{equation}
\mathrm{Hom}(\homo{1}{Y},u(1)^{16}) /
\text{identifications}\ =\ (Y^*)^{16}/\Gamma
\end{equation}
where $Y^*=\BR^5/\Lambda^*$ is the dual torus to $Y$, obtained by modding
out $\BR^5$ by the lattice dual to $\Lambda$. Here we note that $\Gamma$ is
the group of automorphisms of the root lattice of $SO(32)$ (which is also
the group of automorphisms of the $Spin(32)/\BZ_2$ weight lattice).
This agrees with
\eqref{eq:TwistedMod}, provided we identify the M-theory torus $X$, not with
the heterotic torus $Y$, but with its dual torus $Y^*$.

The fact the $X$ is to be identified with $Y^*$ is crucial, as well, to
understanding how the untwisted moduli of the two theories map onto each
other. In trading the torus for the dual torus, we naturally exchange the
3-form $C\in\coho{3}{X}$ of M-theory with the 2-form $B\in\coho{2}{Y}$ of the
heterotic string. Modulo a slight subtlety, which we will encounter in
section \ref{sec:untwisted},  to do with relating the overall volume of
$X$ to that of $Y$, we now understand the mapping between the heterotic and
M-theory moduli spaces.

With this description of the M-theory moduli space in hand, we can examine the
M-theory description of enhanced symmetry groups (which, upon
compactification down to 5 dimensions, become gauge groups) arising in the
limit as some strings become tensionless.

\subsection{The twisted sector}

Consider $n$ 5-branes in the bulk. There is a $U(1)^n$ symmetry, carried by
the ASD 2-forms in the
$n$ tensor multiplets carried by the $n$ 5-branes. Let these $n$ 5-branes
approach each other in the bulk. We need to tune $5(n-1)$ real parameters to
do this. We develop an $U(n)\sim SU(n)\times U(1)$ symmetry. The $U(1)$ is the
``center of   mass" tensor multiplet. Open 2-branes which stretch between
pairs of 5-branes give rise to strings which become tensionless in this
limit. These strings are charged with respect to the tensor multiplets
carried by the respective 5-branes.  The  2-brane which
stretches between the $i^{th}$ and $j^{th}$ 5-brane has charge
$(0,\dots, 1_i, 0,\dots, -1_j,0,\dots 0)$, or minus this, depending on its
orientation.  The corresponding strings are thus in 1-1 correspondence with
the roots of $U(n)$. This point, where $n$ 5-branes coincide is an  $S_n$
orbifold point in the moduli space $\CM$, as we see from
\eqref{eq:TwistedMod},\eqref{eq:GammaGen}. This is to be expected, as
$S_n$ is the Weyl group of the $A_{n-1}$ root lattice.

Now let this collection of $n$ 5-branes approach one  of the fixed points.
This requires tuning $5$ more real parameters. In the limit,
$U(n)$ is promoted to $SO(2n)$. The new tensionless strings come
from 2-branes which stretch between two five branes, passing through the fixed
point on the way. Since the resulting strings effectively change orientation
when they pass through the fixed point, these strings have charge vectors of
the form $(0,\dots, 1_i, 0,\dots, 1_j,0,\dots 0)$, or minus this. Together
with the previous strings, these form the roots of $SO(2n)$. This point in
the moduli space is fixed not just by permutations of the positions of the
$n$ 5-branes, but also by reflections of those positions through the origin.
Thus  the symmetry group is $S_n\semidir (\BZ_2)^{n}$, the automorphism group
of $D_n$\footnote{For $n\neq4$. The full automorphism group
of $D_4$ is $S_3\semidir(S_4\semidir (\BZ_2)^3)$, but only part of this is a
subgroup of $\Gamma$.}.

No surprise, the groups which arise by allowing the 5-branes to move about in
this way are exactly those that arise when tuning the Wilson lines in the
$Spin(32)/\BZ_2$ heterotic string.

\subsection{The untwisted sector}\label{sec:untwisted}

To see other enhanced symmetry groups, we need to tune
also the moduli in the untwisted sector. 
The simplest thing we would like to see is the $SU(2)$ symmetry which arises
when we take one of the radii of the torus $Y$ to the self-dual
radius\footnote{This is $R^2=\alpha'=2M_s^{-2}$ in the
\emph{string metric}. In the 11-dimensional \emph{Einstein} metric, it is
$R^2= \frac{1}{2}M_{pl}^{-2}$, roughly because $X$ is the dual
torus to $Y$.}.

One subtlety, which we now encounter, is that the mapping 
$\bigl(\mathrm{vol}(Y),\ex{\varphi_H}\bigr)\to\bigl(\mathrm{vol}(X),r\bigr)$
mixes these variables in a nontrivial way. This is easily seen, for
instance, from \eqref{eq:hetcoupling}. The $\mpd{5}$ that appears there
differs from the  11-dimensional $M_{pl}$ by a factor proportional to
$(r\ \mathrm{vol}(X))^{1/3}$.

We can realize the T-duality of the heterotic theory on $Y$  as a symmetry of
the 6-dimensional M-theory compactification (that is, without transforming
$r$) \emph{provided} we make compensating changes in the other radii so as
to leave the overall volume fixed. In the case at hand, the relevant choice
is to fix the volume of $X$ to be
\begin{equation}
\mathrm{vol}(X)=\left(\frac{2\pi}{M_{pl}}\right)^5\quad .
\end{equation}

The strings that arise in the untwisted sector consist of 5-branes
wrapped around 4-cycles (which we will call \textbf{A}-strings) on $X$ and
membranes wrapped around 1-cycles\footnote{A 1-cycle is odd under the
$\BZ_2$ symmetry, but so is the 3-form gauge field which couples to the
membrane. So these membranes survive the orbifold projection.}
 (which we will call \textbf{B}-strings). The name, \textbf{B}-string, is
apt. Integrating the 3-form around \emph{one of} the 1-cycles on $X$,
we obtain a 2-form, which is to be identified with the $B$ field of the
type IIB string in 6 dimensions. The \textbf{B}-string associated to this
1-cycle couples to this 2-form and \emph{is} the fundamental IIB string.

Consider the \textbf{B}-string associated to the 1-cycle, $\gamma$, on
$X$. It coupled to a certain 2-form in 6 dimensions, which we might call
$B=(B^-+B^+)$.
As in section \ref{sec:fundstring}, a candidate tensionless string
arises as the bound state of this \textbf{B}-string with a string which
couples to the dual 2-form $(B^--B^+)$. The 5-brane couples to the dual of
the 3-form, so the natural candidate for the string we are after is an
\textbf{A}-string wrapped around the 4-cycle $\gamma^*$ dual to $\gamma$.

With
vanishing moduli of the 3-form $C$, the tension of the \textbf{B}-string
formed by wrapping the membrane around a circle of radius $R$ is 
\begin{equation}
2\pi R T^{(2)}= \frac{M_{pl}^3 R}{4\pi}
\end{equation}
and the tension of the \textbf{A}-string  wrapped around the dual $T^4$ is
\begin{equation}
\mathrm{vol}(\gamma^*) T^{(5)}=\frac{(2\pi)^4}{M_{pl}^5 R}T^{(5)}
=\frac{M_{pl}}{8\pi R}
\end{equation}
where $T^{(2)}$ and $T^{(5)}$ are the membrane and 5-brane tensions.

When the radius of the circle, $\gamma$,
is $\frac{1}{\sqrt{2}}M_{pl}^{-1}$,  the tensions of the \textbf{A}-string and
the \textbf{B}-string are equal and of $\CO(M_{pl}^2)$, but their bound state 
becomes tensionless. This is the string which we studied in section 
\ref{sec:fundstring}\footnote{To make contact with \eqref{eq:twostrings},
note that $(\mpd{6})^4=\frac{1}{2}\left(\mathrm{vol}(X)
\Bigl(\frac{M_{pl}}{2\pi}\right)^5\Bigr) M_{pl}^4$, where the $\frac{1}{2}$
is due to the fact that the volume of the orbifold is half the volume of
$X$.}. In the heterotic picture, the \textbf{B}-string is a mode of the
heterotic string with momentum around the cycle corresponding to $\gamma$.
The \textbf{A}-string is a heterotic string which \emph{winds} around the
same cycle.
T-duality exchanges momenta and windings, and so exchanges these two strings. 
Hence we see that the T-duality
of the heterotic string is an ``electric-magnetic" duality of M-theory,
which exchanges the membrane with the 5-brane! 

Aside from things like the BPS tension (which is unaffected by quantum 
corrections), it is hard to study these strings directly, given our current
rudimentary knowledge of M-theory. The relevant radii of the torus are
$\CO(1)$, and we are far from the regime where low-energy 11-dimensional
supergravity is valid. Nevertheless, it is important to pursue the matter, as
we wish to learn whatever we can about the behaviour of M-theory beyond the
realm of validity of the 11-dimensional supergravity approximation. 

For instance, by further tuning the locations of the 5-branes from the twisted 
sector, we obtain further enhancements of the symmetry group (up to $SO(34)$
in this case). The locations in the moduli space where these and other
enhanced symmetry groups ({\it e.g.}, $E_8\times E_8$) occur are all
well-understood on the heterotic side. We hope to provide a fuller account of
what these look like in the M-theory picture in a future work.


\renewcommand{\baselinestretch}{1} \normalsize

\bibliography{strings}
\bibliographystyle{utphys}

\end{document}